\documentclass[reprint,aps,prl]{revtex4-2}

\usepackage{amsmath,amssymb,amsfonts,dsfont}
\usepackage{graphicx}
\usepackage{color}
\usepackage{hyperref}
\usepackage{subfigure}
\usepackage{dcolumn}
\usepackage{bm}
\usepackage{float}
\usepackage{mathtools}
\usepackage{amsthm}
\usepackage{bbm}
\usepackage{pxfonts}
\usepackage{pstricks}
\usepackage{verbatim}
\usepackage[normalem]{ulem}

\definecolor{orange}{cmyk}{0,0.5,1,0}
\definecolor{rossoCP3}{cmyk}{0,.88,.77,.40}
\definecolor{graa}{rgb}{0.8,0.8,0.8}
\definecolor{blaa}{rgb}{0.2,0.2,0.6}
\hypersetup{
	colorlinks, 
	bookmarksopen, 
	bookmarksnumbered,
	citecolor=blaa, 		
	linkcolor=rossoCP3,	
	urlcolor=rossoCP3,			
	    }

\newcommand{\beq}{\begin{eqnarray}}
\newcommand{\eeq}{\end{eqnarray}}

\newcommand\jt[1]{{\color{purple} [{\bf #1}]}}

\newcommand\gc[1]{{\color{magenta} [{\bf #1}]}}

\newcommand\add[1]{{\color{blue} #1}}

\newcommand{\SU}{\mathrm{SU}}

\newcommand{\U}{\mathrm{U}}

\begin{document}
\noindent\mbox{}\hfill{\small
IPPP/24/58}\par\vspace{2ex}
\title{
Hiding in Plain Sight, the electroweak $\eta_W$}

\author{Giacomo Cacciapaglia}
\email{cacciapa@lpthe.jussieu.fr}
\affiliation{Laboratoire de Physique Theorique et Hautes Energies {\color{rossoCP3}LPTHE}, UMR 7589, Sorbonne Universit\'e \& CNRS, 4 place Jussieu, 75252 Paris Cedex 05, France.}
\author{Francesco Sannino}
\email{sannino@qtc.sdu.dk}
\affiliation{{\color{rossoCP3}{$\hbar$}QTC} \& the Danish Institute for Advanced Study {\color{rossoCP3}\rm{Danish IAS}},  University of Southern Denmark, Campusvej 55, DK-5230 Odense M, Denmark;}
\affiliation{Dept. of Physics E. Pancini, Universit\`a di Napoli Federico II, via Cintia, 80126 Napoli, Italy;}
\affiliation{INFN sezione di Napoli, via Cintia, 80126 Napoli, Italy}
\affiliation{Scuola Superiore Meridionale, Largo S. Marcellino, 10, 80138 Napoli, Italy,}
\author{Jessica Turner}
\email{jessica.turner@durham.ac.uk}
\affiliation{Institute for Particle Physics Phenomenology, Durham University, South Road, DH1 3LE,
Durham, United Kingdom}

\begin{abstract}
The presence of a topological susceptibility in the electroweak sector of the Standard Model implies the existence of a pseudoscalar state in the spectrum, $\eta_W$. We show that, within the Standard Model, no new particle is required to form this state. We identify the $\eta_W$ state with the CP-odd linear combination of the ground states of hydrogen and antihydrogen atoms.
\end{abstract}

\maketitle

\section{Introduction}
Non-Abelian Yang-Mills gauge theories feature a non-trivial vacuum structure, characterised by a non-vanishing topological susceptibility. For $\SU(N)$ gauge theories, there exists a topological $\theta$--term in the Lagrangian,
\begin{equation} \label{eq:topoLYM}
\mathcal{L}_\text{YM} \supset 
\theta\  \epsilon^{\mu\nu\rho\sigma} \text{Tr}\, F_{\mu\nu} F_{\rho\sigma}\,, 
\end{equation}
where  $F^{\mu\nu}$ is the gauge field strength and the trace runs over the $N^2-1$ generators.
When adding fermions non-trivially transforming under the gauge group,  the $\theta$--term becomes related to an anomalous global $\U(1)$ symmetry \cite{tHooft:1976rip}. The presence of this term affects the vacuum and dynamics of the theory with observable consequences via CP-violating effects \cite{Baluni:1978rf}.

In quantum chromodynamics (QCD), the above underlies the strong CP problem \cite{Peccei:2006as}. In fact, the absence of measurable CP violating effects, namely in the electric dipole moments of nucleons \cite{Crewther:1979pi}, implies that $\theta$ is  extremely small. The generalization to different matter representations, multiple strongly coupled sectors relevant for extensions of the Standard Model as well as the proper treatment of the effective action for Nucleons can be found in \cite{DiVecchia:2013swa}. 
A time-honoured, elegant solution to the strong CP-problem would have required at least one massless quark. In this case, $\theta$ would have been unphysical as it could have been removed via an axial rotation. Experimentally, this solution is no longer viable. Even in the absence of quark masses, 
there would still remain a flavour singlet pseudoscalar state, $\eta^\prime$, with a mass generated by the axial anomaly.
For massive quarks, $\theta$ becomes observable and its smallness would require a fine tuning between the QCD vacuum and the phase of the quark masses~\footnote{However, it was recently argued in \cite{Strocchi:2024tis} that the two naturally self-tune to give effectively zero $\theta$.}. Another solution, first proposed by Peccei and Quinn (PQ), \cite{Peccei:1977hh,Peccei:1977ur} consists in introducing a new $\U(1)_{PQ}$ under which the quarks are charged. This allows to rotate away the $\theta$ angle (as long as the quark masses do not break the PQ symmetry). The price to pay is the introduction of a light pseudoscalar, the axion, which emerges from a scalar field that spontaneously breaks $\U(1)_{PQ}$. In the context of QCD, therefore, the Peccei-Quinn $\U(1)_{PQ}$ appears as a \emph{good-quality} symmetry -- leading to a light state -- while the axial $\U(1)_A$ is \emph{bad-quality} -- being broken explicitly by quark masses.

The same topological vacuum structure arises in the electroweak (EW) sector via the $\SU(2)_L$ gauge symmetry, under which left-handed quarks and leptons transform as doublets. In this case, the associated $\theta$ can be rotated away by the anomalous combination of baryon $B$ and lepton $L$ numbers \cite{Anselm:1992yz,Anselm:1993uj}, $B+L$, which is preserved by both gauge and Yukawa interactions~\footnote{This is exactly true for Dirac neutrinos, where the mass comes entirely from a Yukawa coupling to the Higgs. For Majorana neutrino masses, a tiny breaking of $L$ occurs via the dimension-5 Weinberg operator \cite{Weinberg:1979sa}.}.

It has recently been argued that the presence of a topological susceptibility of the vacuum always requires the presence of a pseudoscalar state in the spectrum of the theory \cite{Dvali:2005an}. The argument is based on the fact that the anomalous global Abelian symmetry can be reformulated in terms of a Higgsed three-form gauge field, which contains one massive degree of freedom. Henceforth, this leads to postulate the existence of a pseudoscalar degree of freedom associated to the $\SU(2)_L$ topological anomaly in the SM. This state has been dubbed $\eta_W$ \cite{Dvali:2024zpc}. In this letter, we aim to clarify the nature of such a state and identify a potential candidate within the SM.

\section{The Electroweak Topological Susceptibility}
We first review how the presence of a massive degree of freedom can be inferred from the topological susceptibility of a gauge theory \cite{Dvali:2005an,Dvali:2024zpc}. It is well known that the topological anomaly term in Eq.~\eqref{eq:topoLYM} can be expressed in terms of a three-form \cite{DiVecchia:1980yfw,Hebecker:2019vyf}:
\begin{equation}
    F\tilde{F} \equiv \epsilon^{\mu\nu\rho\sigma}\ \mbox{Tr} F_{\mu\nu} F_{\rho\sigma} = \epsilon^{\mu\nu\rho\sigma} \partial_\mu C^{\rm CS}_{\nu\rho\sigma}\,,
\end{equation}
where the Chern-Simons form is defined as
\begin{equation}
    C^{\rm CS}_{\nu\rho\sigma} = \mbox{Tr}\ \left(A_{[\nu} \partial_\rho A_{\sigma]} + \frac{2}{3} A_{[\nu} A_\rho A_{\sigma]} \right)\,,
\end{equation}
where $A_\mu$ is the 
 gauge field. In a theory where $\theta$ is physical, the correlator of two $F\tilde{F}$ operators must be non-vanishing. Hence the time-ordered ($T$) Fourier-transformed ($F$) correlator of the Chern-Simons form
\begin{equation}
    FT \langle C^{\rm CS},C^{\rm CS} \rangle = \frac{\rho(0)}{p^2} + \dots 
\end{equation}
must develop a non-physical pole at $p^2=0$ (where the dots represent other massive poles). In a theory where the $\theta$ angle is unphysical, e.g. a theory with a good-quality anomalous $\U(1)$ symmetry, the pole must be removed. This could be achieved by decoupling, i.e. $\rho(0) \to 0$, however at the price of inconsistencies with gravity \cite{Hebecker:2019vyf}. The remaining option requires that a mass gap is generated, hence the three form must contain a propagating pseudoscalar degree of freedom \cite{Dvali:2005an}. Note that this conclusion is independent on the mechanism that renders the $\theta$ angle unphysical. In QCD, this degree of freedom can be materialised in the axion \cite{Dvali:2013cpa,Dvali:2018dce,Kaplan:2025bgy}. For solutions not involving the axion, it is the $\eta'$ (e.g. \cite{Nelson:1983zb,Barr:1984qx,Craig:2020bnv,Feruglio:2023uof,Feruglio:2024ytl,Strocchi:2024tis}) or a new composite state (e.g. \cite{Choi:1985cb,Hsu:2004mf,Csaki:2025ikr}).

When we consider the $\SU(2)_L$ gauge symmetry of the SM, the corresponding $\theta$ angle must be rotated away via a phase of the doublet fermion fields, namely the quarks $q$ and the leptons $l$. The SM has a good quality $\U(1)$ anomalous symmetry in $B+L$, which is preserved by all gauge and Yukawa couplings. This $U(1)_{B+L}$ acts on quarks $\mathfrak{q} = (q, u, d)$ and leptons $\mathfrak{l} = (l, e)$ as
\begin{equation}
    \mathfrak{q} \to e^{\alpha/3} \mathfrak{q}\,, \quad \mathfrak{l} \to e^{\alpha} \mathfrak{l}\,,
\end{equation}
where $q$ ($l$) are the left-handed quarks (leptons) and $u,d$ and $e$ denote their right-handed counterparts.
Majorana neutrino masses would break $L$, however the phase could still be removed by use of $B$ (a conserved combination $B-L$ remains). Hence, the $\U(1)$ symmetry remains of good quality as long as $B$ and $L$ are not broken simultaneously, in which case proton decay would occur. The state associated with the $\eta_W$ would, therefore, be generated by the appropriate 't Hooft determinant \cite{tHooft:1976snw} of the quark and lepton fields. To guarantee invariance under $\SU(3)_c$ colour and hypercharge symmetries, the 't Hooft determinant can only be constructed in terms of four-fermion operators, $|\mbox{det} (qqql)|$, where the determinant acts on the fermion family indices. Reference \cite{Dvali:2024zpc} provided a simplified computation, which did not account for colour confinement.
While such a toy model provides an illuminating picture, it does not clarify the physical nature of the $\eta_W$ within the SM.

\section{The Hydrogen Atom: a $\eta_W$ Candidate
}
In analogy to the $\eta'$ in 
QCD, the $\eta_W$ arises as a singlet, not only of the EW $\SU(2)_L$ gauge symmetry but also of the remaining gauge symmetries.
 {The $\eta'$ is sourced by an operator that is a singlet of the flavour symmetry left unbroken by the quark condensate and that carries $\U(1)_A$ charge. The main difference between QCD and the case of the weak $\SU(2)_L$ is that the electroweak gauge symmetry remains weakly coupled due to the Higgs mechanism. Hence, the global flavour symmetry acting on the $\SU(2)_L$ doublets remains unbroken. If we call $\chi^a$ the collection of $N_D$ doublets, the only operator that is singlet under the global $\SU(N_D)$ symmetry and of the gauged $\SU(2)_L$ is
\begin{equation} \label{eq:OW}
    \mathcal{O}_W = \epsilon_{a_1,\dots a_{N_D}} (\chi^{a_1}\dots \chi^{a_{N_D}})\,,
\end{equation}
where $\epsilon$ is the $N_D$-dimensional Levi-Civita symbol.
As we expect the state $\eta_W$ to emerge in the deep infra-red, we can omit the heavy generations of SM fermions and only include the lightest one. Henceforth, in the SM, $N_D=4$ with $\chi^a = (q^r, q^b, q^g, l)$, where $r,b,g$ label the QCD colour indices. The operator in Eq.\eqref{eq:OW}, therefore, reads
\begin{equation}
    \mathcal{O}_W = (qqql)\,.
\end{equation}
This operator has $B+L=2$ (and zero conserved $B-L$), hence it plays the role of saturating the $B+L$ anomaly in the deep infra-red. 
To identify the state sourced by the above operator, it is crucial to highlight the main difference with respect to the QCD case:}
\begin{itemize}
    \item[1)] The EW symmetry $\SU(2)_L \times \U(1)_Y$ is broken to the electromagnetic $\U(1)_{\rm em}$, which is Coulombian in nature.
    \item[2)] Quarks and leptons acquire masses via Yukawa interactions, which preserve $B+L$.
\end{itemize}
Note that the spontaneous breaking of the gauge symmetries does not change the topological structure of the theory, as the symmetry is still present at any scale. Hence, one would still expect a state to emerge in the spectrum, which can play the role of the $\eta_W$. As the $\SU(2)_L$ gauge symmetry is Higgsed, and the coupling is small, the electroweak instantons are suppressed. 
At low energies, the $\eta_W$ must be a QED bound state. Furthermore, only the first generation fermions contribute significantly and the light quarks are confined into hadrons.

Since this state is a QED bound state that only carries $B+L$ charges (but zero conserved $B-L$), an obvious candidate is a bound state of one proton with one electron \footnote{We note that there may be other neutral states such as $n$ $\nu$ but these are not QED bound states.}, i.e. a state of the hydrogen atom whose wavefunction is given by
\begin{equation}
    \Psi_H^{x,s} = \varphi_x (r,\theta)\ \xi_{ep}^s\,,
\end{equation}
\begin{eqnarray} \label{eq:etaWSM}
    \eta_W  &=& \frac{-i}{\sqrt{2}} \left(\Psi^{1s,0}_{H} - \Psi^{1s,0}_{\overline{H}}\right)\,, \\
    \phi_H &=& \frac{1}{\sqrt{2}} \left(\Psi^{1s,0}_{H} + \Psi^{1s,0}_{\overline{H}}\right)\,.
\end{eqnarray}
The former has zero intrinsic angular momentum and being odd under CP, is endowed with the properties of the $\eta_W$. It is, in fact the lightest state within the SM sourced by the operator $\mathcal{O}_W$.

In some extensions of the SM, such as grand-unification models, a mixing between hydrogen and anti-hydrogen is generated \cite{Mohapatra:1982aj}, hence creating a situation familiar in meson systems such as $K_0$--$\overline{K}_0$ mixing -- without CP violation. The $\eta_W$ in Eq.~\eqref{eq:etaWSM}, and the corresponding CP-even combination $\varphi_W$, would be well-defined and observable energy eigenstates. The hydrogen-antihydrogen oscillations \cite{Grossman:2018rdg} would, therefore, be a sign of the misalignment between the energy eigenstates and the states with well-defined baryon number, which are produced in our Universe at the time of the generation of the baryon asymmetry.

\section*{Conclusions and Outlook}
We revisited the implication of the electroweak topological susceptibility for the Standard Model spectrum. Using the three-form/Higgsing perspective, a non-decoupling susceptibility necessitates a pseudoscalar degree of freedom that screens the corresponding Chern--Simons three-form. We showed that no new particle beyond the Standard Model is required: in the deep infrared the role of the $\eta_W$ is played by the CP-odd superposition of the \(1s\) spin-singlet ground states of hydrogen and anti-hydrogen,
\begin{equation}
\eta_W =-\frac{i}{\sqrt{2}}\!\left(\Psi^{1s,0}_{H}-\Psi^{1s,0}_{\bar H}\right)\,,
\quad
\phi_H = \frac{1}{\sqrt{2}}\!\left(\Psi^{1s,0}_{H}+\Psi^{1s,0}_{\bar H}\right)\,,
\end{equation}
and there is also a CP-even combination, $\phi_H$.
This composite state is sourced by the electroweak 't~Hooft operator \(O_W\!\sim\!(qqql)\) for one light generation. Because \(SU(2)_L\) is Higgsed and electroweak instanton effects are exponentially suppressed at zero temperature, the relevant interpolating field, once QCD has confined, matches onto a QED bound state of a proton and an electron at low energies rather than an electroweak meson. 

In the renormalisable Standard Model (or with Dirac neutrinos), \(B\!-\!L\) is conserved and $\Delta(B+L)=4$ transitions are forbidden, so \(H\)–\(\bar H\) do not mix. In extensions with \(B\!+\!L\) violation \cite{Mohapatra:1982aj,Grossman:2018rdg}, operators can be constructed inducing hydrogen–antihydrogen oscillations, in which case \(\eta_W\) and \(\phi_H\) become the  eigenstates, providing a clean experimental handle that directly probes the topological sector via atomic physics.

In the future, it would be valuable to (i) compute the matching \(\langle 0|O_W|\eta_H\rangle\) in a controlled EFT, quantifying how the three-form mass is saturated by atomic states; (ii) estimate \(H\)–\(\bar H\) mixing amplitudes in concrete \(B\!+\!L\)–breaking models and confront existing bounds; and (iii) revisit  constraints on oscillations. 
These steps would sharpen the connection between the electroweak topological sector and low-energy observables, while underlining that the electroweak \(\eta_W\) is an emergent SM composite in the far IR, not a new fundamental particle.

\subsection*{Note added}
In private correspondence, the authors of a forthcoming work by Dvali–Kobakhidze–Sakhelashvili kindly shared with us a forthcoming manuscript from which we learned the following.  They propose realising $\eta_W$ as the pseudo Goldstone of a $B{+}L$–violating 't~Hooft–determinant condensate.  While we agree on the topological necessity of a single pseudoscalar, in their work, the \emph{identity} of $\eta_W$ after QCD confinement is not established, but it is postulated the existence of a new degree of freedom acquiring mass only from the electroweak instantons.  In our work we identify the IR realisation of $\eta_W$ explicitly with the CP-odd hydrogen/antihydrogen bound-state combination, requiring no new particle,  providing a concrete matching, and where the electroweak instantons only provide a tiny contribution to its mass.  

\bibliography{biblio}

\end{document}